# Pulse excitation mode selection via AI Pipeline to Fully Automate the WUCT System

Ankur Kumar[1, 2] and Mayank Goswami[1, #]

[1]*Divyadrishti Laboratory, Department of Physics, IIT Roorkee, Haridwar, Uttarakhand, India*
[2]*Eyepod, University of California, Davis, CA, USA*
[#]mayank.goswami@ph.iitr.ac.in

## Abstract

The parametric optimization for the ultrasound computed tomography (UCT) system is introduced. It is hypothesized that the pulse characteristic directly affects the information present in the reconstructed profile. The ultrasound excitation modes based on pulse-width modifications are studied to estimate the effect on reconstruction quality. Studies show that the pulse width affects the response of the transducer and, thus, the reconstruction. The ultrasound scanning parameters, mainly pulse width, are assessed and optimally set by an Artificial Intelligence (AI)-driven process, according to the object without the requirement of a-priori information. The optimization study uses a novel intelligent object placement (IOP) procedure to ensure repeatability of the same region of interest (ROI), a key requirement to minimize the error. Further, Kanpur Theorem (KT)-1 is implemented to evaluate the quality of the acquired projection data and discard inferior quality data. Scanning results corresponding to homogeneous and heterogeneous phantoms are presented. The image processing step involves deep learning model evaluating the dice coefficient for estimating the reconstruction quality if prior information about the inner profile is known or a classical error estimate otherwise. The model's segmentation accuracy is 95.72% and intersection over union score is 0.8842 on the validation dataset. The article also provides valuable insights about the development and low-level control of the system.

## 1. Introduction

With the advancement of computational tools and emerging new techniques, the development of ultrasound computed tomography (UCT) methods and applications has evolved. In non-destructive testing (NDT), ultrasound has been used for decades as a promising modality to provide accurate and precise detection of defects, material analysis, etc. [1]. Ultrasound computed tomography collectively provides both quantitative and qualitative information. This information can be used to extract 2D/3D volumetric inner profile, relative density distribution, size, and position of the different materials or defects present in the scanned specimen[2, 3]. UCT system extracts the projection data from the attenuated wave (reflected/transmitted) using ultrasonic time of flight or amplitude analysis involving intensive signal processing algorithms [4]. The projection data is then used to reconstruct the 2D/3D cross-sectional profile, solving the inverse problem via iterative or transform methods [5].

The operation of a UCT system involves setting a myriad of parameters. Their fine-tuning and calibration can deliver optimal signal-to-noise ratios, resolution, and contrast [6, 7]. The intricate intertwining of various parameters necessitates a systematic approach to fine-tune these variables. Thus, we introduce parametric optimizations to delve deeper into the intricacies of ultrasound for the enhancement of UCT technology. The objective is to set the scanning parameters automatically for any specimen of interest to achieve better reconstruction results with minimum error in recovery. By systematically optimizing the parameters and their integration for automation, studies aim to minimize human intervention, improve imaging fidelity, and advance the overall UCT technology.

In the literature, parametric optimization has been used for various fields such as X-ray CT, additive manufacturing, optimizing reconstruction techniques, etc. [8-11]. The deep reinforcement learning is used to optimize the parameters for iterative reconstruction to generate the tomograph with fewer projection to minimize radiation dose [10]. A computer-aided training method for hyperparameter optimization in total variation reconstruction in X-ray Computed Tomography is proposed [9]. However, such studies are absent in the context of advancing the UCT systems.

However, to test and implement such a study, control over most of the UCT parameters is required. The presently available 3D UCT come with a fixed set of parameters, which doesn't provide enough flexibility to study the effect of various parameters on the reconstruction [12-14]. The available phased array based 3D UCT system demands extensive electronics associated with each transducer that also adds complexity to the operation, specifically to the production of wavefront. With high production cost and system complexity, phased arrays are not an attractive solution for flexible iterative development [15]. Thus, to study the effect of parameters for optimizing the UCT system, a system simple and flexible enough is required. In this pursuit, a 3D water-coupled system that can provide control over the pulse excitation, number of emitters/receivers, operating voltage, etc., is developed. It requires relatively less extensive electronics but sophisticated control mechanisms and hardware for translation. It eases out the complexities of beamforming. Further, exchangeable transducers and single-channel data acquisition significantly reduce the cost, impart flexibility into system architecture/design, and facilitate easy control suitable for iterative development.

In this article, the parametric studies for (i) optimal resolution, (ii) optimizing the input pulse width, and (iii) setting the optimal input pulse amplitude are presented. The pulse characteristic is one of the main parameters to excite a piezoelectric material, which in turn produces ultrasound wave [16]. In [17], Xiao et al. presented the theoretical framework showing the effect of pulse excitation on transducer response. The effect of driving pulse shapes on transmitted acoustic energy is presented by Choi et al. [18]. Generally, a square pulse accompanied by a rise time and fall time is used to trigger the transducers [19]. It is observed that the pulse characteristics directly affect the transducer response. Apart from the amplitude, the width of the triggering pulse is one of the main parameters affecting the response to the piezoelectric element, as described in [19] by Krautkramer. However, to the best of our knowledge, the effect of pulse characteristics on reconstruction quality has not been studied in the literature. The present article mainly focuses on the pulse-based excitation modes and their effect on reconstruction. In addition, optimization of the number of virtual detectors and applied voltage at the emitter end are also presented.

The classical and AI-based algorithms in pipeline mode are tested to optimize the system[14]. Intelligent object placement algorithms and scanning time optimization are developed to classically optimize performance. Kanpur Theorem-1 (KT1) theorem is implemented to validate the quality of generated projection data. The scanning results of two phantoms: (i) a homogeneous resin phantom (called Multi-sample) consisting of multiple cylindrical rods and cuboidal structures and (ii) a heterogeneous rubber phantom (called Rubber3M) having unknown inner structure and density distribution are presented. To evaluate the similarity of the reconstructed profile, a highly accurate U-Net architecture-based semantic segmentation model is trained and employed to segment the reconstructed profiles. Sørensen–Dice coefficient (DSC) and RMSE error are chosen to assess the quality of the reconstructed profile.

### Motivation
The article focuses on the incorporation of optimization studies for the enhancement of the UCT technology. It is hypothesized that it will reduce the operators' biasedness for operational parameter selection, thus increasing reliability in recovered results. To the best of our knowledge, the research presented in the article has not been explored in the literature. Thus, we seek to fill this gap by providing and implementing optimization studies for the UCT system. Further, to test and implement studies, the design and development of a fully automated and optimized CT system is presented.

Due to unknown heterogeneity in the acoustic impedance distribution of the sample, optimal setting may not be ensured. The parameters need to be set via iterative variation. A strategy must be devised to automate the task without requiring prior information about the object. Steps of automation require optimization of factors such as automatically: (a) adjusting the object at optimal or same location in a pipeline scanning or if its size is dynamically changing after each scan, (b) optimizing the voltage and ensuring the optimal pulse characteristics at the emitter end, (c) number of detectors or the required resolution, (d) estimating the optimality conditions for the above-mentioned operating parameters and (e) mechanism to self-assess the quality of recovered information and discard if the data is of poor quality. The methodology is explained, and results are discussed in forthcoming sections.

## 2. Material and Method

### 2.1. The 3D WUCT System design

A table-top 3D water-coupled ultrasound computed tomography (WUCT) system is developed using a single pair of ultrasound transducers. The hardware assembly is designed to accommodate the electromechanical components and transducers attached to the water bath, all developed in-house. The micro-controller-based electromechanical control system is integrated to precisely control the movement of a sample in 3 spatial coordinates, providing complete 3D scanning. Further, the instrument interfacing, advanced data processing, and image processing algorithms are developed and synchronized to fully automate the system using MATLAB [4, 20, 21]. The graphical user interface (GUI) provides real-time analysis of the scanning process and complete information about the scanned specimen with a single push of a button. The system is designed to provide application specific tuning of various scanning parameters, manually as well as code controlled. The user is given the flexibility to create a *virtual* scanning array consisting of $n$ number of transducers. Since a single transducer-based UCT system requires longer operation time, the current controlling characteristic of the designed circuitry is set to provide longer operation time and robustness without heating up the system electronics. Also, to reduce the scanning time for a single transducer pair based UCT, the scanning speed is optimized by providing the tailored time delay to the control unit according to the chosen scanning parameters.

The designed 3D WUCT system comprises two immersion ultrasound transducer pairs, an arbitrary wave generator (AWG), a digital storage oscilloscope (DSO), an electromechanical control assembly, and a computer. Its block diagram is shown in Fig. 1(a). The perspective view of the scanner assembly is shown in Fig. 1(b). The real-life photo of the working assembly is shown in Fig. 1(c). Movie B shows it in action. The AWG is used to produce a pulse wave to trigger the emitting transducer (Emitter), which in turn produces the ultrasound wave. The generated ultrasound pulse wave gets transmitted via water through the sample. It is detected by the receiving transducer (Receiver) situated on the other side. It is connected to the DSO. The CT algorithm utilizes parallel beam geometry to model the scanning.

### 2.2. Electromechanical Motion Control System

The structural hardware assembly consists of the rotating table, linear and vertical translation platforms coupled with linear ball bearings and threaded-rod systems. It houses the actuators and provides 3 degrees of freedom to the specimen to be scanned. The circuitry is designed to provide a spatial resolution of upto $1.25\ \mu m$ along x, z-axes, and a rotational resolution of upto $0.06°$. An acrylic water container of dimensions $25.5cm \times 13.5cm \times 12cm$ is designed to accommodate the emitter and receiver aligned at $180^0$ angle facing each other. An epoxy resin-based waterproof glue is used to seal and glue the walls of the container. The front face side of the transducers is kept submerged in water ,and the back side is kept outside of the container, which is provided with a UHF connector. The cost efficient NEMA17 stepper motors providing a torque of 6.5 $Nm$ with a step angle of 1.8° are used as the actuator. The actuators are controlled via H-bridge circuitry-based motor drivers, which are connected to a cost-efficient microcontroller Arduino UNO R3. The Arduino is configured to the system through a UART (Universal Asynchronous Receiver Transmitter) serial communication interface via a USB connection [22, 23]. It takes instructions from high-level language (here, MATLAB), decodes them, and performs the low-level control of the actuator motion in real time. Further, micro-stepping circuitry is designed to control motion with higher resolution. As a single transducer-based WUCT takes relatively more scanning time, it needs to be operated for a much longer duration to provide complete 3D scanning, which may result in overheating and wear-tear of the actuators. To prevent any such case, the current flowing through the stepper motor is kept limited to provide just enough torque for the continuous operation.

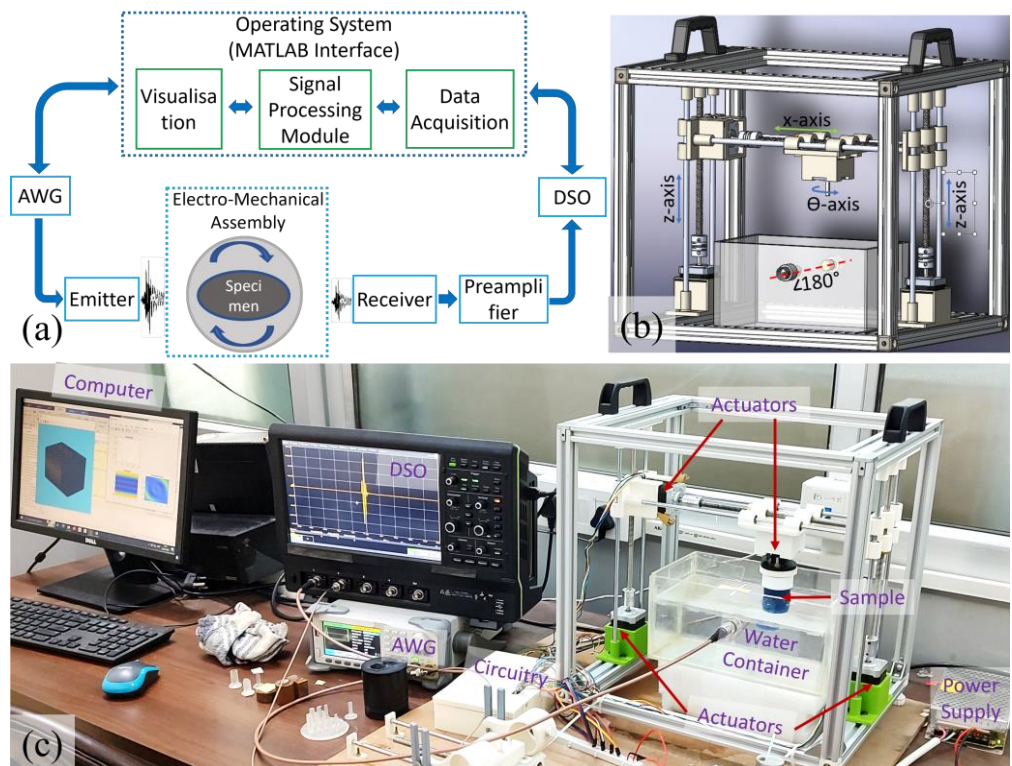

**Fig. 1.** 3D WUCT system, (a) block diagram, (b) CAD model of assembly design and (c) physically realized system. (Note: a video of WUCT system design is provided at Movie A: https://youtu.be/rjlqWdrti9U and Movie B: https://youtu.be/MGnJY7ljn7I?si=G_cmcYBl7bcxNFaI shows it in running state.)

## 2.3. Scanning Time optimization

The major concern is to design the system robust enough to optimize the scanning speed automatically as per the selected scanning parameters. The single transducer pair based system provides several flexibility and low-level control, which comes at the cost of increased scanning time. The scanning parameters, such as the number of linear translations (virtually imitating the number of detectors), number of projections, and number of vertical translations (deciding the number of 2D slices), directly affect the total scanning time and spatial resolution of the final output. The actuators are translated in steps in a particular direction according to the defined scanning parameters; after each step, a delay is provided. This delay time, if not provided efficiently, will result in spoiled reconstruction and cost time. A large delay time will result in a multi-fold increase in the scanning time, while a small delay time will result in the under-ride of the movement, leading to missing translation steps directly affecting the reconstruction. Thus, a relation is derived to estimate a suitable delay considering the step size/distance traveled, actuator speed, acquisition & processing time. A direct correlation between these parameters can be established as:

$$DT \geq C \times \frac{1.25 \times StepSize}{Speed} - \min(t_{ap}) \quad (1)$$

Where, , $DT$ is delay time provided to the control system to induce object movement, $C$ is a constant that can be chosen in the range $1-2$ depending on the application, $StepSize$ corresponds to the number of stepper motor steps executed in a single movement, $Speed$ corresponds to the speed of the stepper motor and $t_{ap}$ corresponds to the time taken in data acquisition and processing.

The relation is implemented to optimize the scanning time implicitly during the experiment. The step will further optimize the operational age of the transducer and prevent the deterioration of the active components of system hardware.

### 2.4. Processing

It includes raw data processing, signal processing, generating graphics, visualizing the processed data, and performing several read/write operations. Data processing mainly involves the algorithms to process the ultrasound signal to extract meaningful data called projection data, and more details can be found in the reference [4, 21]. Projection data is then used to reconstruct the object's profile using a filtered back projection-based reconstruction algorithm. In a single scan, a total of 100 times data is transmitted ($T_n$) for a 100 projection ($P_n$) to reconstruct the profile with $100 \times 100$ resolution. For a single transmission, data is being acquired and processed on board. The amount of data being acquired in a single acquisition is controlled automatically and ranges between 5000-50000 data points ($D_n$) depending on the selected scanning parameters. Thus, for a single scan, a total of $T_n \times P_n \times D_n$ samples are generated. As the data acquired is 12 bits precision, that sums to 24 bits size for both time and amplitude data, which after conversion from byte form to numeric form takes around 43 bytes. Thus, a total of $T_n \times P_n \times D_n \times 43$ bytes data is generated and processed in a single scan. For 3D scan, another term, number of 2D slices ($S_n$) multiplies with the equation to further take a form $S_n \times T_n \times P_n \times D_n \times 43$. The equation provides an estimate of the storage needed before starting the scanning.

### 2.5. Excitation of Ultrasound Transducer

The "Ultran WS37-2" ultrasound transducers built with Lead Meta-niobate piezoelectric material for high bandwidth are used. They have a central frequency of 2 MHz and an active diameter of 9.5mm. The transducers are submersible in water and provided with a UHF connector. On the face, transducers have a quarter wavelength plate designed specifically to allow the efficient transfer of energy from crystals to the water. Transducers can be triggered by an input pulse produced by the AWG to generate the ultrasound wave, whose characteristic mainly depends on the nature of the applied pulse. The pulse width, amplitude, rise edge time and fall edge time of the applied pulse are the controlling parameters that directly affect the nature of the generated ultrasound wave for a specific transducer. The intensity of the generated ultrasound wave is directly proportional to the input pulse amplitude. For a typical transducer, the frequency response has a central frequency for which the transducer is most sensitive. The nature of excitation of piezoelectric material results in variation of the generated ultrasound wave properties [19]. In this section, the effect of pulse-width on the nature of generated ultrasound wave is studied.

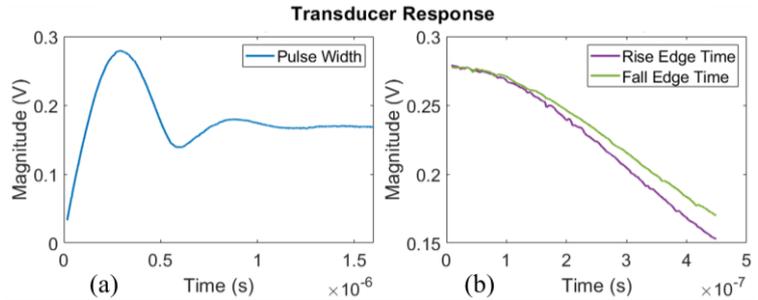

**Fig. 2.** Response of the WS37-2 transducer (a) with change in the applied pulse width, (b) with corresponding change in the rise edge time and fall edge time of the applied pulse.

The response of the ultrasound transducer with respect to changes in pulse width characteristics is shown in Fig. 2(a). The amplitude peaks at the pulse wave of pulse width ~290 *ns* with rise edge time 10.4 *ns* and fall edge time 8.4 *ns*. The transducer response with the change in pulse's rise edge time and fall edge time for a fixed pulse width is shown in Fig. 2(b). This time the pulse width is kept constant at the peak value. The plots show the variation in the nature of generated ultrasound wave and the transduction efficiency of the transducer when the pulse-width of triggering pulse wave is changed. Its effect on the reconstruction is studied in Section III.

A typical input pulse having rise and fall edge time of 10ns having an amplitude of 20V is applied to trigger the transducer. The generated ultrasound wave signal corresponding to a single input pulse is shown in Fig. 3(a). An increase in the pulse width (keeping rise/fall edge time constant) results in a change in the frequency and amplitude of the generated ultrasound wave, as shown in Fig. 3(b) & (c).

The acoustic attenuation changes with the change in generated ultrasound wave characteristics, for the fractal microstructures of media, such acoustic attenuation typically exhibits a frequency dependency characterized by a power law [24-27], given by eq.:

$$P(x + \Delta x) = P(x)e^{-\alpha(\omega)\Delta x} \qquad (2)$$

$$\alpha(\omega) = \alpha_o|\omega|^y, \quad y \in [0,2] \qquad (3)$$

Where, $p(x)$ and $p(x + \Delta x)$ are the acoustic pressure at point $x$ and $(x + \Delta x)$, respectively. The $x$, $\omega$ and $\alpha$ represent the displacement, angular frequency, and the attenuation coefficient, respectively. Here, $\alpha_o$ and y are constant depending on the material properties and sound wave frequency.

It can be understood from the eq. 3 that a change in $\omega$ will induce $\alpha(\omega)$ to change. In simple words, pulse wave used to trigger ultrasound transducer will have a direct effect on the $\omega$ and hence on $\alpha(\omega)$ which in turn impact the wave attenuation. In other words, pulse characteristic will affect the penetration of the ultrasound wave and the transmission intensity will be different for the different materials.

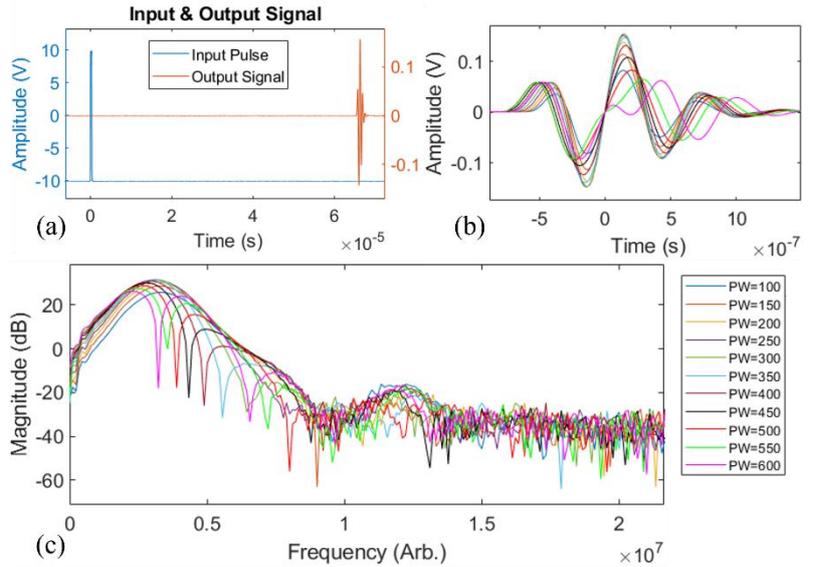

**Fig. 3**: Response of the WS37-2 transducer, (a) input pulse and generated ultrasound (Output) signal, (b) change in nature of the generated ultrasound wave with change in pulse-width values and (c) frequency response for selected PW values.

In general, the ultrasound wave attenuates as they propagate through the object. In this case the object can be modelled as a two-dimensional (or three-dimensional) distribution of the attenuation coefficients and a line integral represents the total attenuation suffered by the ultrasound wave as it travels in a straight line through the object. The projection $P_\theta(t)$ can be represented in relation to the object $f(x,y)$ as [28],

$$P_\theta(t) = \int_{(\theta,t)line} f(x,y)\, ds \qquad (4)$$

So, the direct effect of the pulse characteristic can be seen on the reconstructed profile of the object. Thus, the effect of pulse width change on tomographic reconstruction is studied for homogeneous and heterogeneous samples.

## 2.6. Intelligent object placement (IOP)

One of major sources of error when scanning an object multiple times while changing scanning parameters is object placement. The orientation and the position of the object is not the same when it is placed again for a second scan, third scan…. so on, this induces an inherent inconsistency in the scanning results. It is true for any CT system; the exact results are quite difficult to reproduce. Thus, an automatic object detection and placement algorithm is developed and integrated into the WUCT system. Once the object to be scanned is placed in the sample holder. The scanning parameters, such as the number of 2D slices, views/projections, and detectors, are selected according to the scanned object size and required resolution. The algorithm detects the edge of the object multiple times and then fine-tunes the edge position to exactly locate the position of the object. The algorithm is based on the changes in signal characteristics once the object interferes with the emitted ultrasound wave.

Initially, the object is placed into the holder and translated to the extreme position on the horizontal translation axis ($x$-axis). Then, the object is translated in $x the$ direction in fixed step sizes (say $l$), data is acquired using the same transducers at each increment. The onboard analysis of data provides feedback on the object's location. As soon as the first edge of the object interferes with the ultrasound wave, it results in a change in the signal. Then, moving the object back by 1 step and further incrementing the position in smaller steps ($l/10$) to finetune the edge position. Now, the position of the first edge is located. Then, object position is further incremented in larger steps ($l$) until we see the normal signal without object interference that confirms the position of the second edge. Again, finetuning the second edge position. Process allows to place the object in a known location on the $x$-axis. A similar algorithm is followed to confirm the orientation while rotating the object.

Orientation of the object is important when the object is scanned a second time. In this case, the intensity is matched to that of the earlier orientation for a complete line scan. When the mismatch (a threshold value above which the acquired signal is the same) is minimal, it will confirm that the orientation is the same. Data is then processed to get information about the object's position and start the scanning procedure. This would not have been possible unless the data processing and hardware control algorithms were not integrated into each other. The algorithm ensures minimum human intervention and scanning results reproducibility.

## 2.7. 3D WUCT Operation

Initially, the object to be scanned is placed in the holder. The scanning parameters, such as required resolution, number of projections, pulse characteristic, scanning height/width, option to save raw data, etc., are selected (required resolution is the main parameter; other parameters are automatically selected if not provided). Then, the system fixes the position of the object at the optimal location using the IOP procedure. Then, AWG triggers the transducer with the specified pulse characteristics. In turn, the emitter produces ultrasound waves, which traverse through the object then detected by the receiver. The parallel beam geometry is applied to acquire the scanning data, and filtered back-projection algorithm-based inverse radon transform is used to reconstruct the object's profile [29, 30]. For a single 2D slice, the transmission data is acquired for the input number of detectors (selected according to the resolution and obtained by translating the object at multiple positions) for each projection. The data is processed on board for each acquisition and visualized after each projection. After collecting data for all the detectors, the object is rotated to acquire data for another projection, and the process is repeated until all the data is collected for a single 2D slice. Once data is acquired for one slice, the object position is shifted vertically, and the whole process is repeated to acquire data for the next slice. The process is repeated until data is acquired for all the slices. Then, the reconstructed profile from each slice is stacked to form the 3D volume.

## 2.8. Pulse Width Optimization

The transducer behaviour is different for different PW values, which can be observed from Fig. 2(a). So, it is assumed that the best PW will result in the maximum transmission of ultrasound wave generating CT data faithful to reconstruction modeling via inverse problem solution. The sinogram (projection CT data/line integrals) will thus capture most of the information of the inner profile of the object. Classically, one may simply choose the optimal PW resulting sinogram with the maximum mean pixel value. However, a sinogram with the most information will have most of the features appearing in its image perceivable to the human eye only. Thus, the AI model is trained to select the best PW based on the feature tracking. However, feeding a full sinogram corresponding to different PW values into the trained model would result in a time-consuming process and will not have any practical use as one has already collected data for all PWs. In contrast to the full scan, a single line scan can be acquired in a few minutes. So, a process that can select the best PW based on just one line scan is developed to provide instant selection. To efficiently implement the process, a random forest classifier (RFC) is trained to predict the best PW [31]. It consists of a user-defined number of tree classifiers where each classifier is generated using a random vector sampled independently from the input vector, and each tree casts a unit vote for the most popular class to classify an input vector [32]. The major advantage of using the RFC model is to prevent overfitting on the selected data [33].

For the model training, line scan data corresponding to the 100 different projection angles for all the considered PW values is generated by scanning the two mentioned samples. The line scan data consists of the processed signal value from the data acquired for each of the 100 unique detector positions for a specific projection angle. Line scan corresponding to all PW values for one projection angle case is shown in fig. 4. In total, 11 different PW

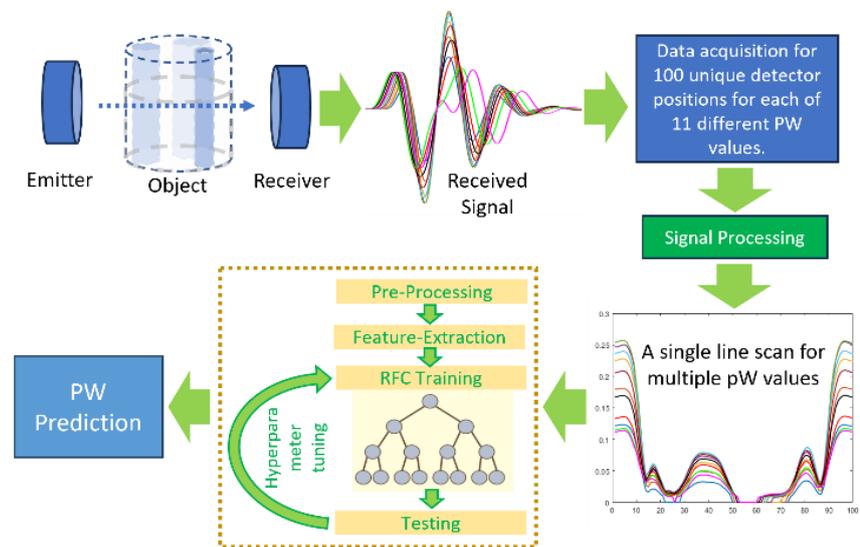

**Fig. 4**: Process diagram for the PW prediction using random forest classifier.

values are chosen in the active range of the transducer. This data for all PW values is pre-processed to find the features affecting the reconstruction. Five key features are observed to change, namely,

i) total number of maxima: It changes depending on the number of discontinuities present in the object. A higher value for a specific PW value suggests that the corresponding ultrasound wave can resolve the discontinuities.
ii) total number of minima: Same effect as 'total number of maxima' and considered for better consistency.
iii) average value: It is observed that the average value should be larger for better contrast.
iv) maximum value: The maximum value in the line scan after removing any outlier (if exists) is directly related to higher penetration through the object and also affects the contrast value. It can be chosen accordingly depending on the interest of the user. In our case, we have chosen it to have larger value for better penetration, thus observing all the possible discontinuities in the object.
v) minimum value: This feature is directly related to ultrasound penetration. A minimum value suggests low penetration and a higher value suggests better penetration. Thus, it was chosen accordingly for the required application. In our case, we have chosen it to have a larger value.

These features are generated for each sample data and fed to the classifier model for training. A detailed diagram showing the process of PW prediction using the RFC model is shown in Fig. 4. The hyperparameters are tuned using GridSearchCV and RandomizedSearchCV techniques. After tuning, the n_estimators=150, criterion='log_loss', 'max_features'=None, 'max_depth'=10, and 'max_leaf_nodes'=6, are selected. The model has shown an accuracy of 75% on the test dataset. The accuracy of the model can be further improved by feeding more data.

Initially, the WUCT scans the object, saving single line scan data for a defined number of unique detector positions, each time varying the PW. These line scan data are fed as input to the trained RFC model to evaluate the best PW for the corresponding object. In response, the model outputs the best PW value. For better accuracy, multiple line scan data at different angles can be fed to the model for an overall best PW (or a combination of PW) value. Afterward, the WUCT scans for a relatively higher number of translations.

## 2.9. AI-based quantitative analysis

An established artificial intelligence-based tool to semantically segment the reconstructed images is implemented for quantitative analysis. Semantic segmentation is the segmentation of the input image according to semantic information and predicts the semantic category of each pixel from a given set [34, 35]. An end-to-end fully convolutional neural network U-Net architecture-based deep learning model is used for the image segmentation [35]. The architecture basically consists of two paths, contraction, and expansion, which combine to form a U-shaped structure. The model is selected as it doesn't require a large amount of training data for better accuracy and works better even when the training data is limited [35]. The generated results are used to evaluate the Sørensen–Dice coefficient (DSC), hence the similarity in the reconstructed and actual phantom image.

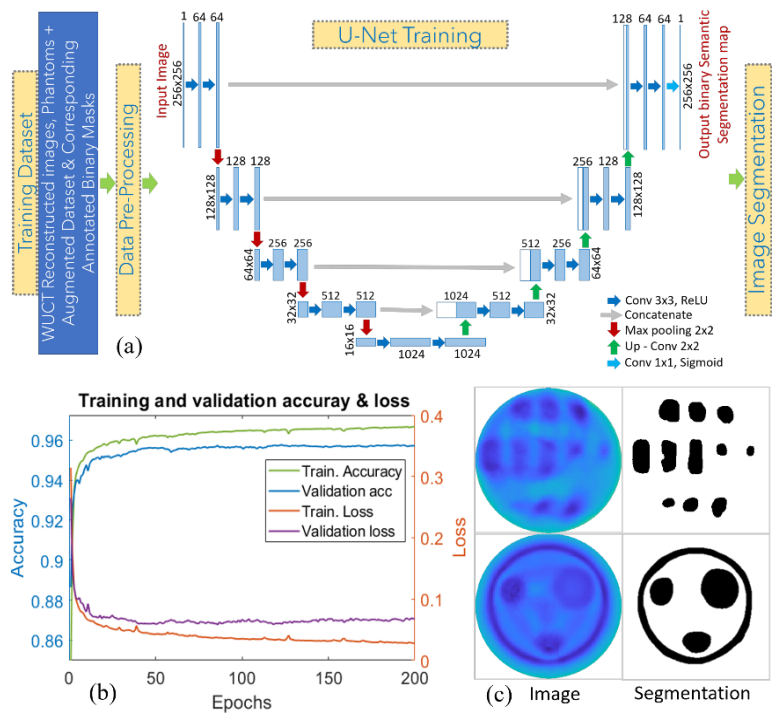

Reconstructed images are pre-processed to form a training dataset. The number of original images was limited, so the training data is augmented by manipulating the images[36, 37]. The model is trained on a dataset of 500 images including WUCT reconstructed, phantom, and augmented images. The augmented image dataset is generated using the *albumentations* library by applying image transformations on the original dataset. Basic manipulation such as rotation, image transposing, horizontal and vertical flipping are used. The dataset is semi-automatically annotated to generate binary masks for the training.

The detailed architecture of the model is shown in Fig. 5(a). Rectified linear unit is used as the activation function for each of the convolution layer [38]. The sigmoid is used as the activation function for the output layer. Stochastic gradient based optimizer 'Adam' is used to adaptively update the weights of the neural network [39]. The binary cross-entropy is used as the loss function to evaluate the discrepancy between predicted and actual values[40]. The model is trained on the labelled dataset for 200 epochs, the training and validation results are shown in Fig. 5(b). The accuracy of the model on the validation set is 95.72 % and the IoU (Intersection over Union) score is 0.8842. The predicted binary segmentation on test images is shown in Fig. 5(c). The equation (Eq. 2) to evaluate DSC for an overlapping pair of phantom and reconstructed images can be written as:

$$DSC = 2 * \frac{N_o}{N_p + N_r} \tag{5}$$

**Fig. 5.** U-Net semantic segmentation model, (a) process diagram for U-Net model training, b) training and validation accuracy and loss, (c) predicted segmentation on the test images.

Where, $N_o$ is the number of overlapping pixels, $N_p$ and $N_r$ are the pixels in the phantom image and reconstructed image, respectively. The original images are replaced by the segmented images to calculate the DSC values.

The dice coefficient ranges between 0 and 1. For an acceptable match, DSC should be close to 1. To evaluate the DSC, two criteria must be met: (i) the reference and image objects must be of similar size, and (ii) these objects must exhibit a high degree of overlap [41].

## 3. Results

### 3.1. Data Quality assessment

The scanned results of two samples to study the achievable resolution and scanning capability are presented. Also, a study to estimate the similarity and quality of the reconstruction profile with change in resolution and pulse-width is carried out. The quality of projection data generated using the WUCT system is validated using the KT-1 theorem on both of the goodness of fit (gof) parameters, namely, RMSE and $R^2$ [42]. KT-1 estimates the inherent overall noise in projection data. It is used to assess the effect of: (i) increase in the number of virtual transducers and (ii) variation in pulse-width on the projection data, as shown in Fig. 6. Both of the gof: RMSE and $R^2$ suggests the projection is of high quality. Fig. 6 shows that this error estimates (irrespective of RMSE or $R^2$) is either less sensitive to conclude which resolution value and PW value are optimal or the effect of these two parameters over reconstructed image quality is small.

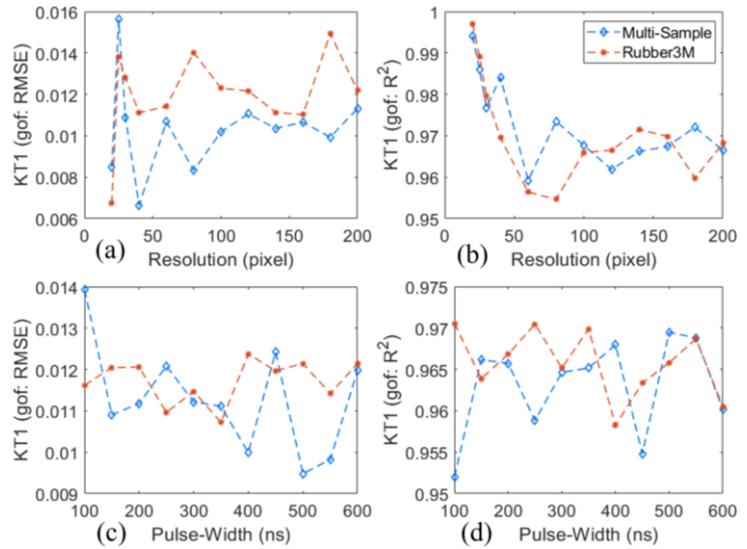

**Fig. 6.** KT-1 analysis of the projection data based on (a) & (c) gof: RMSE and (b) & (d) gof: $R^2$ for data corresponding to (a) & (b) Resolution study and (c) & (d) Pulse-width study.

### 3.2. Parametric optimization

1. **Amplitude optimization:** Input pulse amplitude is responsible for the intensity of the generated ultrasound wave. The voltage at the emitter is usually chosen based on prior information of sample material to ensure just enough wave transmission to the receiver. A sub-optimally lower value may reduce SNR, or a higher value may induce unnecessary loading on the transducers' acoustically active piezoelectric element, especially during long scanning hours [43]. The ultrasound transducer crystals and membranes are susceptible to damage for higher amplitude values of the input pulse. Thus, the effect of variation in the input pulse amplitude on the reconstruction quality is studied to obtain the optimal value. The input pulse voltage is changed between 4-20V range in a step size of 2V. The profile of the heterogeneous object Rubber3M is reconstructed for a 100x100 pixel resolution by creating the virtual array of the source and detector by translating the object in fixed-length steps. The contrast variation, RMSE, and dice similarity coefficient values are evaluated for each of the profiles as shown in fig 7. The contrast in the image is directly proportional to the variation in the intensity of the reconstructed image thus a plot of the intensity distribution is obtained for each of the profiles, as shown in Fig. 7(a). It is observed that higher values of the input pulse amplitude produce better contrast in the reconstructed profile. Also, the RMSE decreases linearly as the pulse amplitude is increased, although the variation in the RMSE is only in a very small range as the object is scanned for a fixed resolution for each of the amplitude values, as shown in Fig. 7(b). The DSC coefficients suggest that the structural similarity reaches ~89 % after an amplitude value of 8V and then further starts to saturate, as shown in Fig. 7(c). The study suggests that the amplitude value directly affects the contrast of the reconstructed profile. This can also be understood from eq. 2. However, the higher structural similarity can be achieved at a smaller value of amplitude. It also provides the estimation of the minimum threshold value of the input pulse amplitude required to be set.

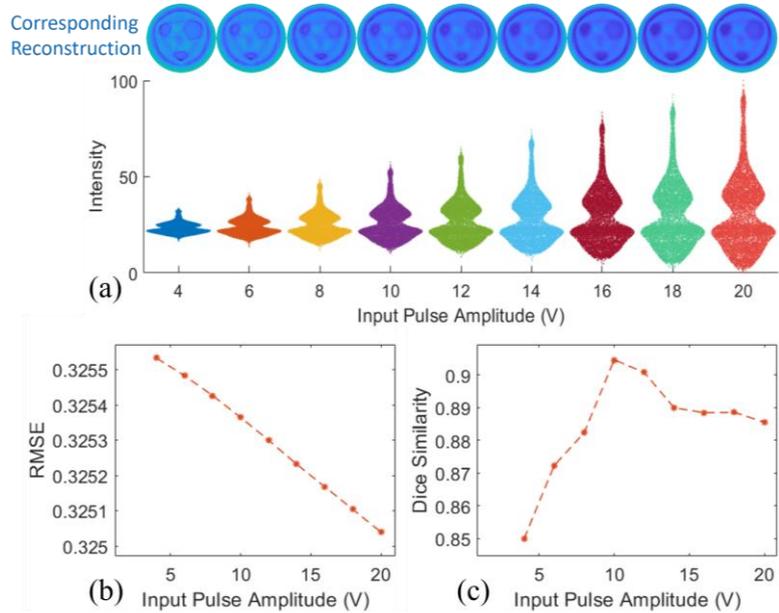

**Fig. 7**. Variation in the reconstructed profile with change in the input pulse amplitude, a) contrast variation and corresponding reconstructed profile, b) variation in RMSE and c) dice similarity coefficients.

2. **Resolution optimization:** The effect of an increase in the number of virtual transducers on reconstruction quality is studied by increasing the scanning resolution in the range of 20x20 to 200x200. Total number of virtual transducers is directly related to the resolution as for a case of 20x20, 20 virtual transducer positions are required. The study is carried out for both the sample, Multi-sample and Rubber3M. Virtual phantom corresponding to each of the resolutions is generated to compare the reconstructed profile for the same resolution as shown in Fig. 8. Reference phantom images corresponding to the Multi-sample and Rubber3M are shown in Figs. 8(a) and 8(c), respectively. While the reconstructed 2D profiles corresponding to the Multi-sample and Rubber3M are shown in Figs. 8(b) and 8(d), respectively. Reconstructed profile similarity and quality is evaluated using DSC and RMSE. Phantom and reconstructed profiles are segmented using the trained semantic segmentation model to generate the binary mask corresponding to both images for each of the resolution data. The evaluated DSC values are found to be in a range of ~0.86 to 0.94. At first, values started to increase with an increase in the resolution, then saturated after a value greater than 60x60 for both samples, as shown in Fig. 9(a). This behaviour suggests that the further increase in the resolution will not significantly improve the similarity in the reconstruction. However, the RMSE error continued to reduce with an increase in the resolution for both samples, although the reduction is not prominent after a resolution of 100x100 as shown in Fig. 9(b). The result suggests that a resolution profile of 100x100 pixels or greater is good enough to scan the selected samples with good accuracy. However, better resolution is required to resolve the finer details. Thus, a 100x100 resolution profile is selected to study the variation in the reconstruction quality with change in pulse-width.

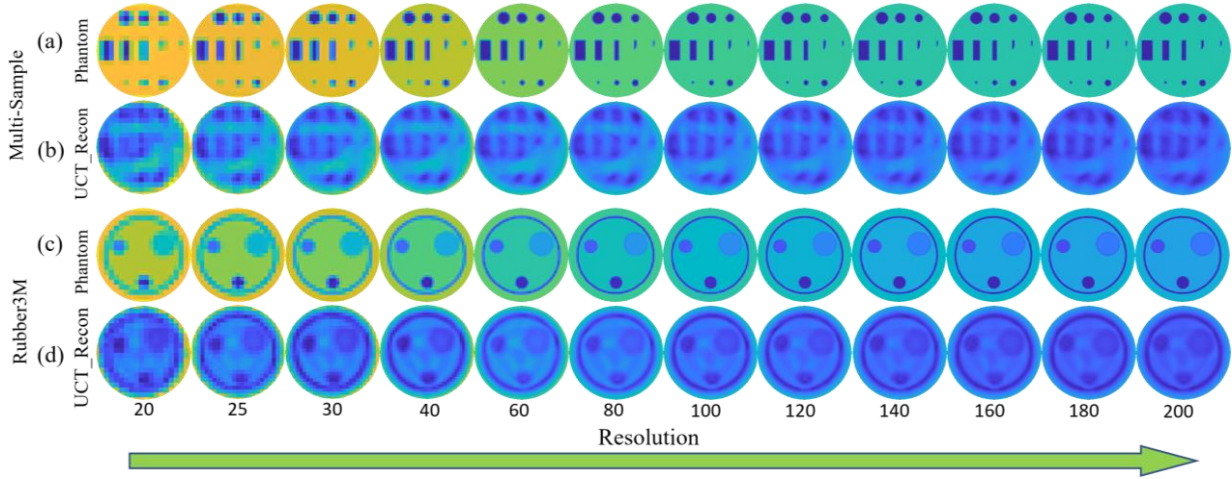

**Fig. 8.** Variation in the reconstructed profile with change in the resolution for Multi-Sample and Rubber3M, (a) and (c) variation in the phantom image, (b) and (d) variation in the reconstructed profile.

3. **Pulse-Width optimization:** The change in pulse characteristics, such as pulse-width, directly affects the nature of the generated ultrasound wave and its amplitude, thus affecting the attenuation. To study the effect, pulse-width corresponding to the range 100ns-600ns is used with an increasing step of 50ns. The reconstruction profile of both the homogeneous and heterogeneous samples is obtained for scanning resolution of 100x100. The acquired data in the form of the sinogram and corresponding reconstructed profiles are shown in Fig. 10.

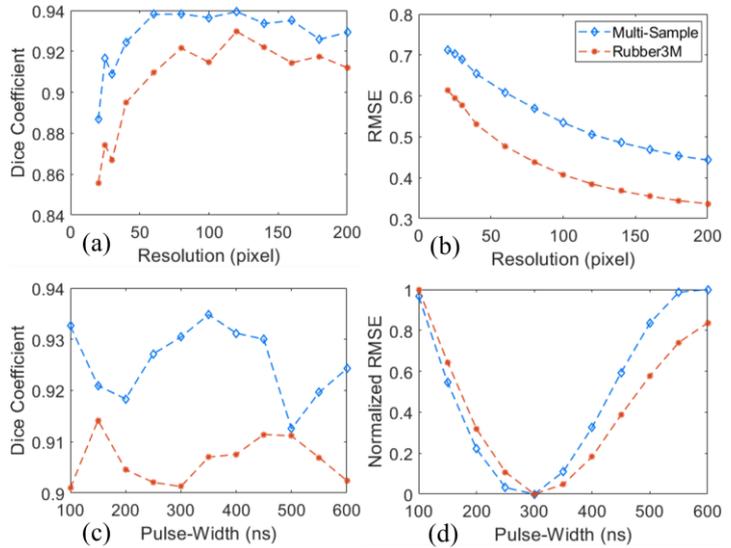

**Fig. 9.** Estimation of the error in the reconstructed profile of Multi-Sample and Rubber3M for (a)&(b) resolution study and (c)&(d) pulse-width study using (a) and (c) dice similarity coefficients, (b) and (d) RMSE.

The curve showing DSC with the change in pulse-width is shown in Fig. 9(c), the range on DSC is found to be in the range of 0.9-0.94. However, the change in DSC value doesn't show any clear trend, suggesting the similarity of the image with the phantom does not change much. For Multi-sample, the change in the structure of the reconstructed profile is almost negligible. However, the reconstruction quality is better for a pulse-width of 300ns, which is also better for the Rubber3M sample supported by RMSE error results, as shown in Fig. 9d). Also, if we look at the sinogram of the Multi-sample (refer to Fig.10(a)), the change in reconstruction is negligible, as shown in Fig. 10(b). But for the heterogeneous Rubber3M sample, the penetration of the ultrasound wave increased with an increase in the pulse-width as observed from the sinograms, as shown in Fig. 10(c). The corresponding reconstructions also show the penetration of ultrasound waves through the rods present in the structure. If we closely look into the reconstructed profiles (Fig. 10(d)) we find an inside circular region in each of the rods whose size is gradually reducing with an increase in the pulse-width, suggesting low attenuation of ultrasound waves (a high-quality video Movie C is provided for a clear comparison).

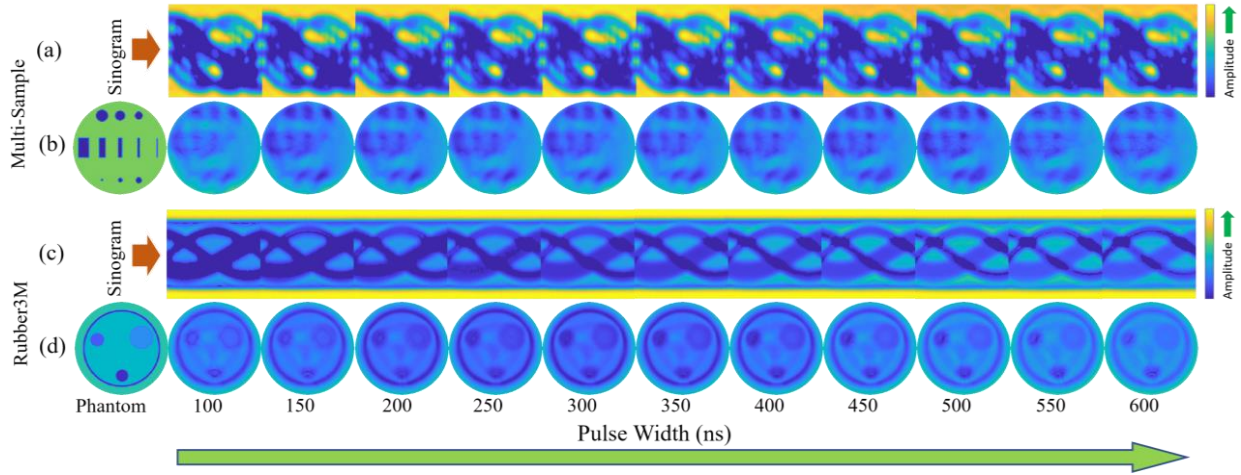

**Fig. 10.** Variation in the reconstructed profile with change in the applied pulse-width for Multi-Sample and Rubber3M, (a) and (c) variation in the sinogram, (b) and (d) variation in the reconstructed profile. (Note: a high-quality video to visualize variation is provided at Movie C: https://youtu.be/CdkIq84oK_Q .)

To select the optimal value of PW for a particular object at a particular scanning angle, an RFC model is trained. The model selects the best PW value based on certain features in the line scan data. The 200 samples corresponding to both samples for the PW study are tested. The results obtained are shown in Fig. 11(a) & (b). The method shows that a certain value of the PW may not be the best for scanning at all the projection angles and a combination of PW values is obtained. The RFC results suggest that a PW value of 300ns is best for 56%, 250ns best for 20%, 350ns is best for 13%, and the rest PW values are best only for 11% of line scans for the Multi-sample. However, for heterogeneous Rubber3M, 250ns, 300ns, and 350ns are best for 31%, 34%, and 27% of the line scans, respectively. This suggests that for a heterogeneous sample multispectral imaging is necessary. The results support multispectral imaging where a single value of frequency value is not sufficient to get the best reconstruction.

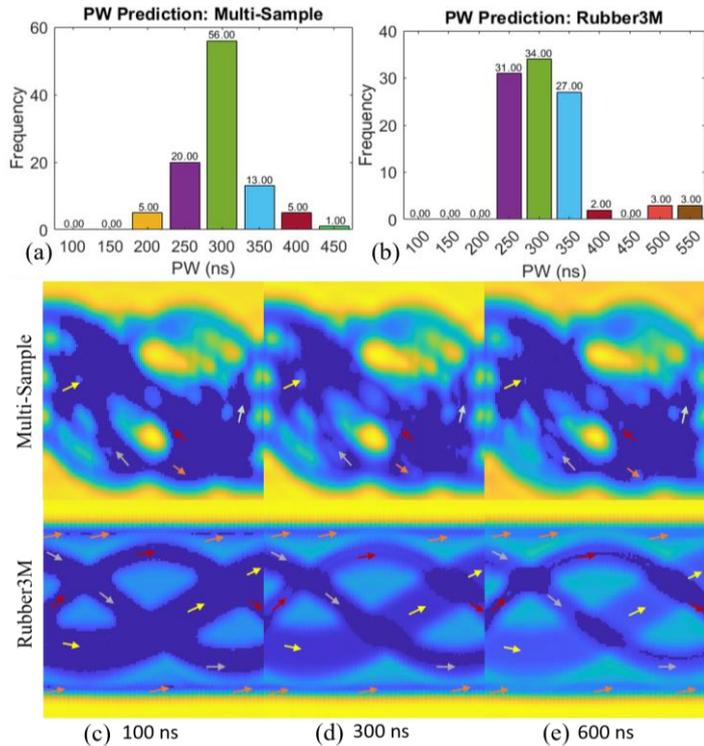

**Fig. 11.** Optimal pulse width prediction, (a) for Multi-Sample, (b) Rubber3M, and feature tracking in the sinogram using different coloured arrows for different features of Multi-sample and Rubber3M for PW (c)100ns, (d)300ns and (e)600ns.

The features are tracked to visually identify feature-rich sinograms among three chosen values as shown in Fig. 11(c), (d) & (e). The different colour arrow marks show the appearance and

Submitted to arXiv on 09th August 2024

disappearance of certain features in the sinogram for both the samples. For example, in case of Multi-sample, a yellow-colored arrow represents that the feature made a larger footprint in the sinogram corresponding to PW of 300ns (Fig. 11(d), for Multi-sample case), then its structure almost disappeared for PW value of 600ns as visible from Fig. 11(e). Similar observations are made for Rubber3M sample, see the red arrow that also shows the appearance and disappearance of the feature. These features are quantified and added to the RFC model for predicting the best PW value.

### 3.3. 3D WUCT reconstruction and analysis

1. **Multi-sample:** A homogeneous 3D sample is designed to check the scanning resolution of the system in the presence of multiple samples, the phantom is named Multi-sample. The 2D CAD design and 3D printed sample made up of resin are shown in Figs. 12(a) and 12(b), respectively. The design consists of six cylindrical structures and five cuboidal structures. Out of the five rectangular structures, two structures of thickness 0.05 cm and 0.1 cm broke due to their weak structure, although some of the portion of these structures remained attached to the sample. The phantom is intentionally chosen to study as it will help in the analysis in a more rigorous way. The Multi-sample is scanned for a profile resolution of $100 \times 100$. A total of 24 slices are obtained at a vertical resolution of 0.1 cm to reconstruct the 3D profile. While scanning the object, an air bubble got trapped at the location of the broken structure, as shown in Fig. 12(c) and the zoomed image is shown in the inset. The Multi-sample is used as it is so that the bubble's profile will also be reconstructed along with it. The WUCT took 66127 s to complete the scanning and reconstruct the 3D profile. The scanning results are shown in Fig. 12(d), all the structures present in the sample are clearly reconstructed in the 3D profile. Also, the trapped bubble's profile is clearly visible with a well-defined boundary (shown in the inset). For better visualization a video (Movie D) is provided. The outside area around the structures is present due to the spreading of the wavefront produced by the transducer. The analysis suggests that the object upto 0.05 mm can be easily resolved with the WUCT using a transducer of 9.5mm active diameter.

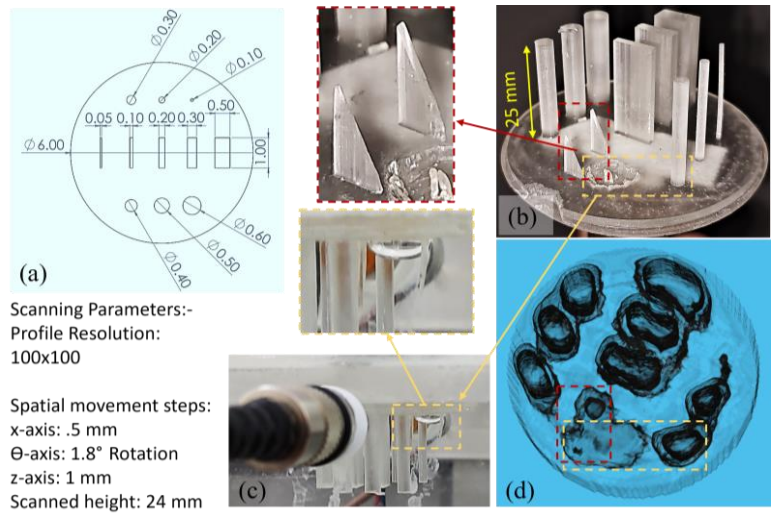

**Fig. 12.** Sample with multiple cylindrical and cuboidal structures (Multi-Sample), (a) 2D drawing with structural dimensions (units in cm), (b) image of the physical sample, (c) a bubble at the surface of the sample while scanning (the zoomed figure is shown in inset), (d) 3D reconstructed profile of the specimen. (Note: a video of 3D reconstruction is provided at Movie D: https://youtu.be/3ZXV7gZ5BHk .)

2. **Rubber3M:** A heterogeneous sample having different material density is scanned for a profile resolution of 40 × 40. A total of 60 slices at 0.1 cm vertical resolution are obtained to reconstruct the 3D profile. The actual image of the sample is shown in Fig. 13(a). A small part of the sample is chipped off from the bottom, which is glued, as shown in Fig. 13(a) (zoomed view is shown in inset). The inner distribution of the sample is obtained by X-ray radiograph as shown in Fig. 13(c) & (f). The scanning results are presented in Figs. 13(b), 13(d) & (e). The WUCT took 31805 s to complete the scanning and reconstruct the 3D profile. The scanned profile clearly shows three rods along with the chipped off part. To highlight the specific portion in the volume, a volume viewer tool is used [44]. Relative variation of the sectional profile is shown in Fig. 13(d). The result suggests that rod #1 has a comparatively higher density than the other two. Gray colour in Fig. 13(d) shows higher penetration of the ultrasound wave because of better impedance match. The observed density order of the rods is $\rho_1 > \rho_2 > \rho_3$, although the difference in the density of the rods is varying only slightly which can be observed from the 2D radiograph of the sample, as shown in Figs. 13(c) & 13(d). The density contrast is also better in WUCT reconstruction showing clear variation in between the rods, and object material as visible from Fig. 13(e). From the 3D profile, it is also found that the height of rods is different, which can be observed from both the results, as shown in Figs. 13(b) & 13(c). The WUCT is also able to reconstruct the profile of the chipped off part with clear estimate of the boundary and volume. The volumetric scanned profile of the chipped off part is shown in the inset of Fig. 13(b) & (e). A video Movie E showing the complete 3D scanning result of sample, and the damaged section is provided for better visualization of the results.

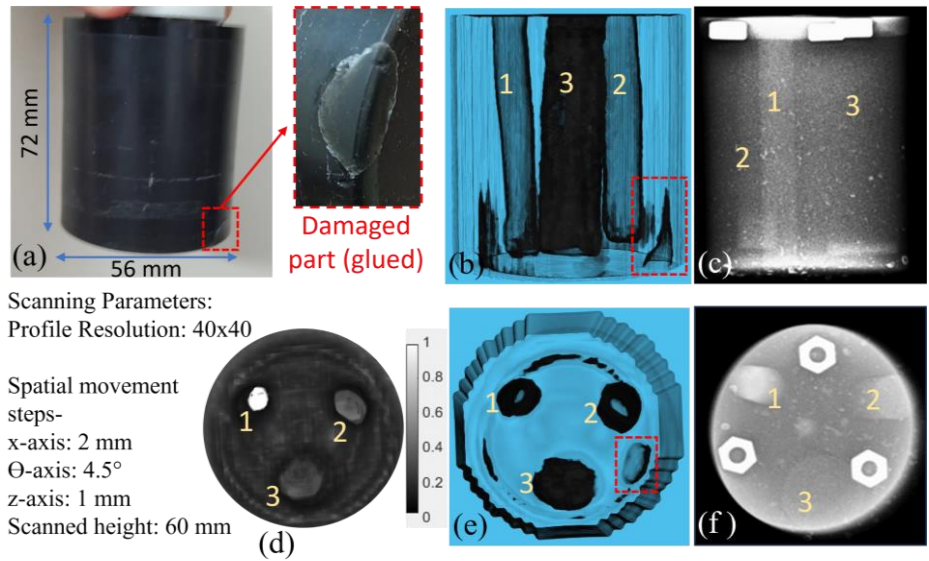

**Fig. 13.** Rubber sample with 3 rods (Rubber3M), (a) sample image with defect (zoomed image of chipped off part of the sample in inset) (b) reconstructed inner profile, (c) x-ray radiograph, (d) relative density distribution of sample as obtained from WUCT scan, and top view of sample profile using e) 3D WUCT, (f) X-ray radiograph. (Note: a video of 3D reconstruction is provided at Movie E: https://youtu.be/rFmsnYxjSH8)

## Discussion:

The common problem with 3D ultrasound CT systems is the generation of large amounts of data. In a single experiment, hundreds of GB of data are generated, resulting in the depletion of storage resources. With the generation of large data, the required processing power and time consumption also increase. This can be minimized by using storage-efficient file formats, saving the raw data when necessary, decimation, compression, etc. Further optimization can be done through controlled data acquisition, which focuses on the data containing the required information.

The minimum resolution that can be achieved by an ultrasound transducer mainly depends on the operational frequency and the size of the transducer aperture. The larger aperture size results in an acquisition of the wavefront from a wider area, limiting the resolving capabilities of the system. In the WUCT system, transducers having a central frequency of ~ 2MHz are used, which could have resolved object sizes up to 0.015 cm. However, the larger aperture size limits the practically achievable resolution.

Assessing the quality of reconstruction images quantitatively is a challenge, as the image does not have similar colors, although the image is similar. In this case, if we employ RMSE, SSIM, etc., methods that depend only on the pixel value are prone to produce errors or show low sensitivity detection when used alone without being supported by any index that quantifies the structural similarity. For example, the location of the defect and structure mismatch from the actual image due to slight rotation will produce an entirely wrong analysis. However, when a structural similarity index is used firsthand, then a pixel-wise quantifying tool is used, and the results are valuable. Thus, we have used an AI segmentation-based DSC coefficient that provides information about the similarity of the structures, not about the quality of similarity. Then, evaluating the RMSE, after the dice coefficients can establish the structure position similarity, provides the measure of the quality of an image.

For the segmentation, the conventional segmentation methods such as active contour are also used [45]. However, the conventional techniques are not adaptive in comparison to deep learning techniques. The results are not good because of the absence of well-defined boundaries in the images of the reconstruction profiles. In the case of conventional techniques, the parameter values set for one type of image may fail when applied to slightly different images. In contrast, the deep learning-based model recognizes the complex patterns in multiple images to produce more accurate predictions.

# Conclusion

The optimization studies are tested and successfully implemented into the developed 3D WUCT system. The parametric optimization studies served as a tool to set optimal parameter settings to improve the performance of the system. Further, classical and AI tools integration fully automated the tested system. To show practical applicability, homogeneous and heterogeneous samples are scanned using the WUCT system.

The scanning time optimization algorithm implicitly optimizes the delay time, improving the operating performance of the system.

Excitation modes of ultrasound transducers based on the pulse characteristics improve the contrast and accuracy of the reconstruction at optimal settings. The study also motivates us to deeply analyse the generated ultrasound wave and its effect on the reconstruction and the ways we can alter the nature of the generated ultrasound pulse for specific applications.

An increase in the number of virtual transducers on reconstruction quality does not vary after an optimal value is obtained. After the optimal value, the error starts to saturate. In the study, it is found that a profile resolution of 60x60 is good enough to scan with the WUCT system.

The amplitude study shows that a lower input pulse amplitude value may also result in an accurate reconstruction profile (similar based on DSC). However, for better image contrast, a higher input amplitude value is desired.

PW variation affects the reconstruction of the homogeneous and heterogeneous samples differently. In the case of the heterogeneous object (Rubber3M), an increase in pulse width resulted in an increase in acoustic penetration, which is also observed in corresponding reconstructed profiles.

An RFC model is trained to implicitly optimize the PW setting according to the object. Prediction results show that a certain value of the PW may not be the best for scanning a particular sample and strongly supports multispectral imaging for heterogeneous samples. The scanning using appropriate PW values would improve the reconstruction quality.

The scanning results of the WUCT for the homogeneous and heterogeneous sample show that-

1. The generated 3D profile is structurally similar to the original 3D objects. For Multi-sample, the structures upto 0.05cm thickness are reconstructed. Also the trapped bubble's reconstructed profile is also clearly visible with a well-defined boundary.
2. The heterogeneous Rubber3M sample scanned profile shows the variation in density as expected. Also, the defect boundary and volume can be estimated from the 3D reconstructed profile. This shows the potential for further exploring the NDT applications for complex heterogenous structures.

The results confirm that the optimization of the scanning parameters improves the similarity and contrast of the reconstructed profile. Also at the same time, it opens a window to selectively enhancing the scan quality to study specific properties or ROI. In the future, a detailed correlation of multiple parameters can be obtained, which will not only improve the scan quality but also improve the application-specific autotuning of the multiple parameters.

In the future, the objective is to develop advanced signal processing techniques considering the non-linearity of the ultrasound propagation to extend the system's applicability to clearly scan metal and complex objects.

## Declarations:

### Acknowledgments:

AK acknowledges CSIR for the fellowship and contingency grant. MG would like to acknowledge the support of Early Career Research Grant code ECR/2017/001432 by the Science and Engineering Research Board (SERB), Government of India.

### Credit authorship contribution statement:

AK: Hardware design and development, Methodology, Data Acquisition, Instrumentation, Investigation, MATLAB & AI codes, Visualization, Writing; MG: Methodology, Funding, Supervision, Writing.

### Conflicts of Interest:

The authors declare the following financial interests/personal relationships, which may be considered potential competing interests: Part of this work is submitted for the patent at the Indian Patent Office with application no. 202311078074.

### Data availability:

The data corresponding to the study shown in the article is made available at https://dx.doi.org/10.21227/ehrj-z518 and the processing codes can be made available on request.